\newif\ifREFEREE \REFEREEfalse 
 
\ifREFEREE  
	\documentclass[referee]{aa501} 
	\usepackage[nomarkers]{endfloat} 
	\AtBeginDelayedFloats{} 
\else 
	\documentclass{aa501} 
\fi 
 
\usepackage{graphicx,amssymb} 
\usepackage{natbib} 
\bibpunct{(}{)}{;}{a}{}{,} 

\newcommand{\mycaption}[1]{ 
	\ifREFEREE    \caption[#1]{} 
	\else        \caption[]{#1} 
	\fi 
	} 
 
\begin{document} 
 
 
\title{H$_2$ formation and excitation in the diffuse interstellar medium} 
 
\titlerunning{H$_2$ formation and excitation in diffuse ISM} 
 
\author{ 
C. Gry\inst{1,2} \and 
F. Boulanger\inst{3} \and 
C. Nehm{\'e} \inst{3} \and 
G. Pineau des For{\^e}ts\inst{3} 
 \and E. Habart\inst{3}
 \and E. Falgarone\inst{4}
 }
\authorrunning{C. Gry et al.} 
 
\institute{ISO Data Center, ESA Research and Scientific Support Department, 
PO Box 50727, 28080 Madrid, Spain \and 
Laboratoire d'Astrophysique de Marseille, BP 8, 13376 Marseille Cedex 12, France
\and
Institut d'Astrophysique Spatiale, Universit{\'e}  Paris Sud, Bat. 121, 
91405 Orsay Cedex, France 
\and Ecole Normale Superieure, Laboratoire de Radioastronomie, 24 rue Lhomond,
75231 Paris Cedex 05, France
}  

\offprints{C{\'e}cile Gry, cgry@iso.vilspa.esa.es} 
             
\date{Received 02 October 2001/ Accepted 06 May 2002} 
 
\abstract{ 
We use far-UV  absorption spectra obtained with FUSE towards
three late B stars  to study the formation and 
excitation of H$_2$ in the diffuse ISM. 
The data interpretation relies on a model of the chemical
and thermal balance in photon-illuminated gas. 
The data constrain well the $n \, R$ product between gas density
and H$_2$ formation rate on dust grains: $n \, R$ =
  1 to $\rm 2.2 \, 10^{-15} s^{-1}$. 
For each line of sight the mean effective H$_2$ density $n$, assumed uniform, 
is obtained by the best fit of the model to the observed N(J=1)/N(J=0) ratio,
since the radiation field is known. Combining $n$
 with the $n \, R$ values, we find  similar H$_2$ formation rates
 for the three stars of about $R = 4 \,10^{-17} $cm$^{3}$s$^{-1}$.\\
 Because the target stars do not interact with  the absorbing matter we
can  show that the H$_2$ excitation in the $J> 2$ levels cannot be 
accounted for by the UV pumping of the cold H$_2$ but implies collisional 
excitation in regions where the gas is much warmer. 
The existence of warm H$_2$ is
corroborated by the fact that the star with the largest 
column density of CH$^+$  has the largest amount of warm H$_2$.
} 
\maketitle

\section{Introduction} 

The H$_2$ formation is a key process for the understanding of 
the thermal and density structure as well as the chemical evolution of the
interstellar medium (ISM). 
The H$_2$ formation rate was first estimated through the modelling
of the hydrogen recombination on dust surfaces (e.g. Hollenbach et al. 1971).  
Based on an analysis of Copernicus observations
of atomic and molecular hydrogen in the local diffuse clouds, 
Jura (1975a) 
proposed an H$_2$ formation rate ($ R = 3 \, 10^{-17} {\rm cm^3 s^{-1}} $)
which  corresponds to the Hollenbach
et al. (1971) prediction for a total grain surface of $\rm 10^{-21} {\rm cm}^2/H$
and a recombination efficiency of 0.5. 

The excitation of the H$_2$ rotational levels from the ground
state observed in absorption in the UV  is a diagnostic of physical conditions. 
In diffuse clouds, the low J lines provide a measure of the gas temperature while the
excitation of the $\rm J>2$ levels 
is generally interpreted as a result of the fluorescence cascade following 
H$_2$ pumping by the UV radiation from the target  stars.
But collisional excitation in shocks driven by the star 
have also been considered.

Since the pioneering work of Black \& Dalgarno (1976) on which the Jura analysis 
of Copernicus data is based, much progress has been made in 
the modelling of H$_2$ in space, in particular about the
fluorescence cascade after UV pumping of electronic transitions and
collisional deexcitation rates (Combes \& Pineau des For{\^e}ts 2000). 
The {\it Far Ultraviolet Spectroscopic Explorer} (FUSE) is  also now providing 
new UV absorption observations of
Galactic H$_2$ superseding the Copernicus observations by their
sensitivity (Snow et al. 2000, Shull et al. 2000, Rachford et al. 2001). 
The topic of H$_2$ formation and excitation has also been revived by the 
observation of the mid-infrared transitions between the  rotational 
levels of the vibrational ground state. These data have been used to estimate the H$_2$ formation rate in warm
photo-dissociation regions at the surface of molecular clouds (Draine \& Bertoldi 1999, 
Habart et al. 2002) where the gas and the dust are both warmer than in the 
diffuse ISM and thus
where the H$_2$ formation efficiency or the processes involved might differ.
ISO observations are not sensitive enough to detect the mid-IR H$_2$ line
 emission in the 
low to moderate column density lines of sight studied in the UV but they have 
allowed to detect an extended warm 
H$_2$ component away from star forming regions across the Galaxy (Verstraete 
et al. 1999) and in the edge-on galaxy NGC 891 (Valentijn \& van der Werf 1999).
The Galactic data has been interpreted as evidence for the existence of warm
H$_2$ gas heated by the dissipation of kinetic turbulent energy.

In this paper, we re-consider the
question of H$_2$ formation and excitation in the diffuse ISM by analysing
FUSE observations of three late B stars located behind the Chamaeleon clouds. 
IRAS images show that these stars unlike most earlier type stars usually observed in the
UV are truely background field stars
that do not interact with the matter responsible for the absorption and that do not contribute
to the incident radiation field. 
These three stars were part of a larger sample of Chamaeleon lines of sight observed with IUE 
to correlate the UV extinction curve with changes in the dust size 
distribution traced by the IRAS mid-IR to far-IR colors (Boulanger, Pr{\'e}vot, Gry 1994).
The interstellar medium along these lines of  sight was further characterized 
with high resolution optical and ultraviolet absorption spectra 
with a spectral
resolution of 10$^5$ obtained at ESO and UV spectra with the
Goddard high resolution spectrograph on the Hubble Space Telescope 
(Gry et al. 1998). Characteristics of these lines of sights and previously 
observed column densities are given in Table~\ref{tab:columns}.

The FUSE data results (Sect.~\ref{sec:h2col}) are analysed
by modelling  the interaction of the gas with the Solar 
Neighborhood radiation field.
This analysis leads on one hand to an estimate of the
H$_2$ formation rate (Sect.~\ref{sec:form}) and on the other hand to 
a discussion of the H$_2$ excitation (Sect.~\ref{sec:warm}) where we 
argue that H$_2$ column densities
at $J> 2$ cannot be acounted for by UV pumping.  
\section{Observations and data analysis}\label{sec:h2col}
High resolution far UV (905--1187 \AA) spectra of HD102065, HD108927 and 
HD96675 have been obtained in April and May 2000 by FUSE observations of
respectively 6.8, 6.5 and 10.2 ks, as part of the PI Team Guaranteed
Time.
FUSE design, data processing and  performance have been described in 
Moos et al. (2000) and Sahnow et al. (2000). The data have been 
reprocessed with the calibration pipeline version 1.8.7.

To derive the H$_2$ column densities, we have used the bands located in
the best part of the spectra, namely 
the Lyman (0,0), (1,0), (2,0), (3,0), (4,0) and (5,0) bands.
In effect the  UV flux of these stars decreases rapidly with wavelength 
shortward of 
Ly$\beta$, due to their relatively late types  and their interstellar 
extinction. The stellar Ly$\beta$ absorption is so broad and deep in these
spectra that it completely wipes away the interstellar line, prohibiting the
determination of HI column densities. 
\begin{figure}[h]
\ifREFEREE 
 \rotatebox{90}{\resizebox{\hsize}{!}{\includegraphics{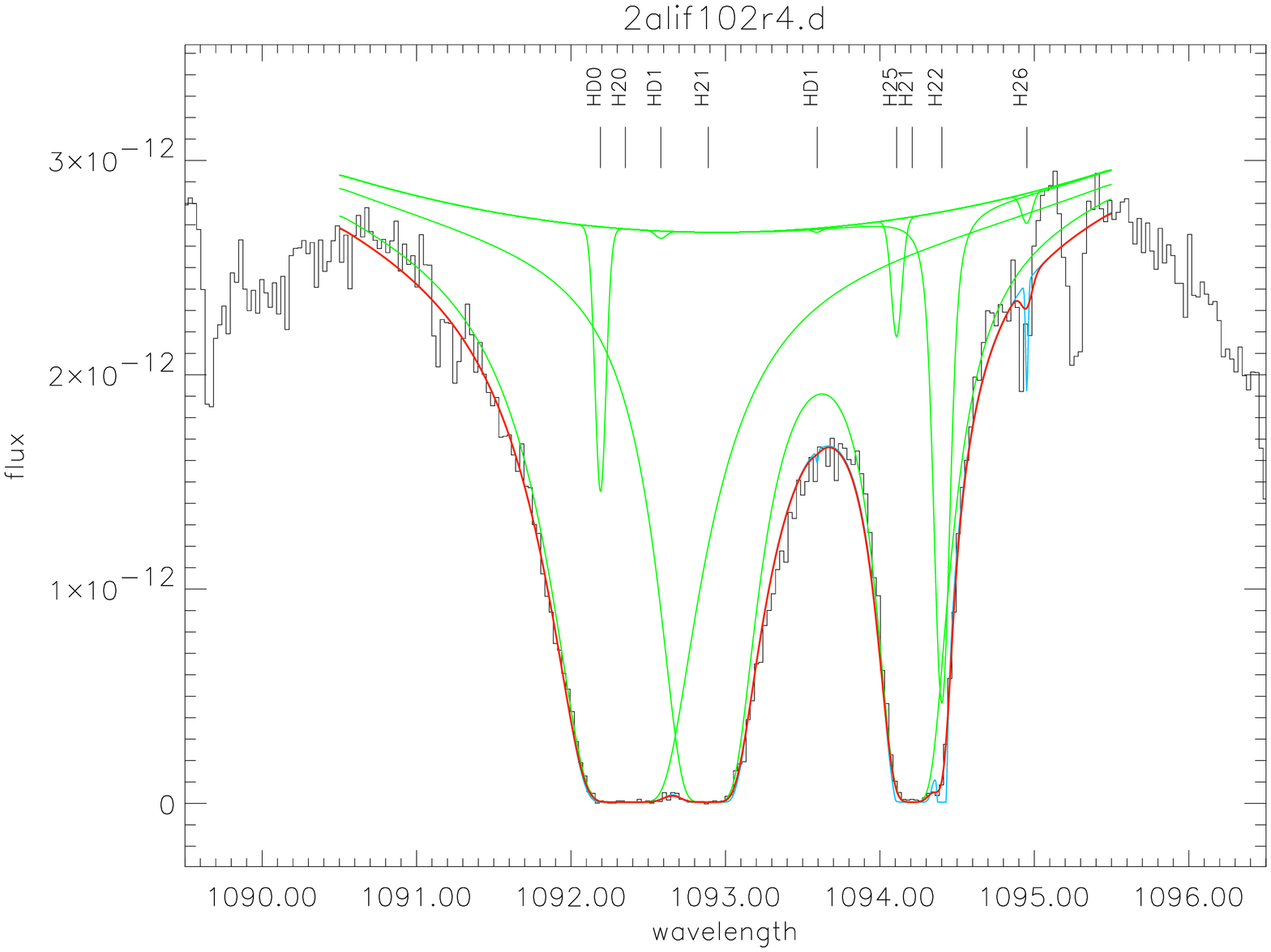}}}
\else
\rotatebox{90}{\resizebox{6.5cm}{!}{\includegraphics*[68,710][534,75]{h3639f1.ps}}}
\fi
\mycaption{Example of  fit of  the H$_2$ J=0 and J=1
level transitions in the line of sight towards HD102065. The plot shows the 
FUSE observations, the synthetic absorption components due
to the individual transitions, and the resulting profile.
Only one band (Lyman(1,0)) is shown but all available
bands are fitted together. Only the column density of the J=0 and J=1
levels (marked H20 and H21) are derived from the fit. The higher H$_2$ levels 
(in this case J=2, J=5 and J=6) as well as the HD lines
are included in the fit, but  their column densities cannot 
be derived from this fit. 
\label{fig:spectrum}} 
\end{figure}

The J=0 and J=1 H$_2$ lines are  heavily saturated and their Lorentzian
wings allowed us to derive precise  column densities for these two levels 
independently of the gas velocity distribution. 
  They have been measured  through  profile fitting 
with the  program ``Owens'' which allows to fit together (blended) 
lines from  different H$_2$ levels  and other species. All available 
bands are fitted together meaning that the derived column density gives
the best fit on all bands together. This allows to decrease the errors 
mainly due to the uncertainty on the continuum placement because of
overlapping wings of  adjacent bands.
The error bars on column densities have been estimated empirically as the 
dispersion of the results of  five  fits performed with different sets of four 
H$_2$ bands each.

The derived H$_2$ column densities are  
 listed in Table~\ref{tab:columns}. 
An example of the data and the fits is shown in Fig.~\ref{fig:spectrum}
for the Lyman(1,0) band in the line of sight towards HD102065.
\begin{table}[h] 
\begin{center}
\leavevmode
\footnotesize
\begin{tabular}{l|c|c|c}
\hline 
    &HD102065 &HD108927  & HD96675\\
\hline
Sp. Type & B9IV & B5V & B7V \\
d(pc) & 170 & 390 & 250 \\
E(B-V) & 0.17 $\pm  $ 0.04 & 0.23 $\pm  $ 0.04 & 0.31 $\pm  $ 0.03 \\ 
A$_{\rm V}$ & 0.67 $\pm  $ 0.12 & 0.68 $\pm  $ 0.10 & 1.1 $\pm  $ 0.15 \\
R$_{\rm V}$ &3.9 $\pm  $ 0.4 & 3.0 $\pm  $ 0.2 & 3.5 $\pm  $ 0.2  \\
N(CH) & 6.3 10$^{12}$ &1.4 10$^{13}$& 2.2 10$^{13}$ \\
b(CH)  &1.9 km s$^{-1}$& 2.2 km s$^{-1}$& 1.6 km s$^{-1}$\\
N(CH$^+$) & 1.2 10$^{13}$ & - &2.8 10$^{12}$ \\
b(CH$^+$) &3.0 km s$^{-1}$& - & 2.2 km s$^{-1}$\\
N(CO) & (7 $\pm  $ 3) 10$^{13}$ & -&$> 10 ^{15}$ \\
N(C\,{\sc i}) & (4 $\pm  $ 1) 10$^{14}$ & - &$> 1.7 \, 10^{15}$ \\
\noalign{\smallskip}
N$_{\rm total}$ & 9.9 10$^{20}$ & 1.3 10$^{21}$ & 1.8 10$^{21}$ \\
\noalign{\smallskip}
\hline 
\noalign{\smallskip}
N(H$_2$,J=0) &$2.0 \pm 0.2\,10^{20}$ &$2.0 \pm 0.2\,10^{20}$  &$ 4.9 \pm 0.9\,10^{20}$\\
N(H$_2$,J=1) &$1.4 \pm 0.1\,10^{20}$ &$1.2 \pm 0.1\,10^{20}$  &$ 2.1 \pm 0.6\,10^{20}$ \\
\noalign{\smallskip}
$f$=$\frac{\rm 2 N({\rm H_2})}{\rm N(H)}$  & 0.69$\pm  0.12$ & 0.49$\pm  0.09$ & 0.76 $\pm  0.15$\\
\noalign{\smallskip}
\hline
\end{tabular}
\mycaption{Interstellar absorption toward the stars.\\ Column densities 
are in cm$^{-2}$. \\
CH and  CH$^+$ column densities were 
derived from high resolution (R=110\,000) optical spectra  obtained with
the ESO 3.6~m telescope and the Coude Echelle Spectrometer. The C\,{\sc i} column 
densities were derived from HST GHRS observations
(Gry et al. 1998).\\ The total column density, N$_{\rm total}$, has been
derived from E(B-V), N$_{\rm total}$ = 5.8 10$^{21}$ E(B-V). 
\label{tab:columns}}
\end{center}
\end{table}

\begin{figure}[h]
\ifREFEREE 
  \resizebox{\hsize}{!}{\includegraphics*[40,340][510,720]{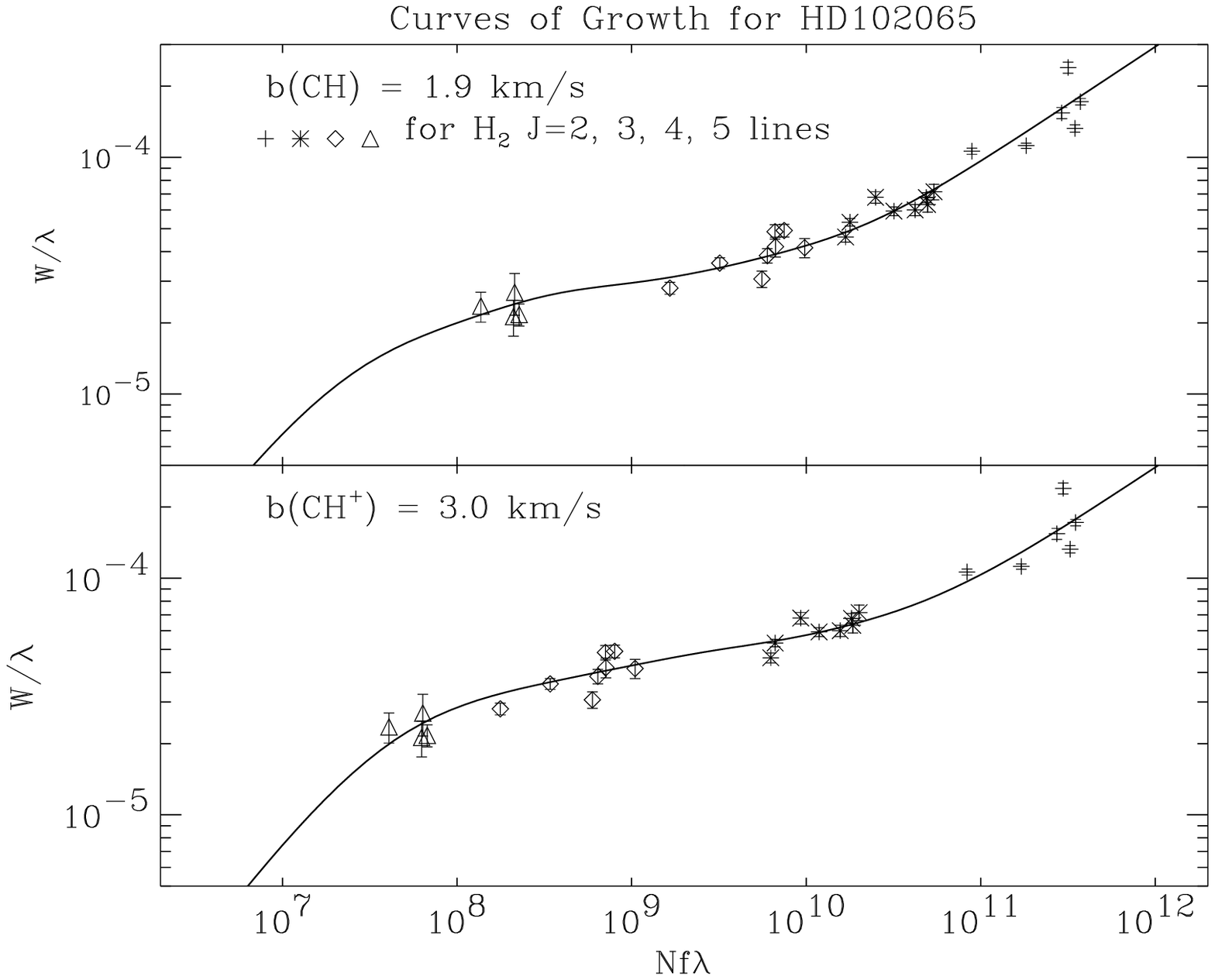}}
\else
\resizebox{!}{7cm}{\includegraphics*[40,340][510,720]{h3639f2.ps}}
\fi
\mycaption{Curves of growth for the excited levels of H$_2$. The top plot 
shows the curve of growth for
b(H$_2$) = b(CH) and the bottom plot that
for b(H$_2$) = b(CH$^+$). The J$>1$ column densities listed in 
Table~\ref{tab:columns}
have been derived by fitting the measurements of all J$>1$ levels together 
to both curves of growth respectively.
\label{fig:growth}}
\end{figure}

For the higher excited levels, the lines are less saturated and
fall on the flat part of the curve of growth where
the column densities depend heavily on the gas velocity distribution.
Thanks to high resolution optical observations of the molecules CH and CH$^+$,
we have derived the b-value of the gas responsible for the CH absorption 
for the three lines of sight and of the gas responsible for the CH$^+$ 
absorption for two of them (Table~\ref{tab:columns}).
For HD102065 we also have a measurement of the   C\,{\sc i} b-value from
new high resolution HST-STIS data: b(C\,{\sc i}) = 1.8 $\pm0.1$ km/s, close to
b(CH).
\begin{table*}
\begin{center}
\leavevmode
\footnotesize
\begin{tabular}{l  cc  c  cc}
\hline 
    &\multicolumn{2}{c}{HD102065} & HD108927  & \multicolumn{2}{c}{HD96675}\\
\hline
b$_{\rm adopted}$&1.9 km/s&3.0 km/s&2.2 km/s&1.6 km/s&2.2 km/s\\
\hline
	\noalign{\smallskip}
N(H$_2$,J=2) &$2.6\,10^{18}$ &$2.5\,10^{18}$& $1.7\,10^{17}$ &$ 3.3\,10^{18}$ &$ 3.4\,10^{18}$\\
N(H$_2$,J=3) &$3.1\,10^{17}$ &$1.1\,10^{17}$&$ 9.3\,10^{16}$ &$ 1.7\,10^{17}$&$ 8.7\,10^{16}$\\
N(H$_2$,J=4) &$5.6\,10^{16}$ &$6.0\,10^{15}$&$ 1.5\,10^{15}$ &$ 4.0\,10^{15}$&$ 1.4\,10^{15}$\\
N(H$_2$,J=5) &$1.2\,10^{15}$&$4.5\,10^{14}$ & $\leq 6\,10^{13}$ & \multicolumn{2}{c}{$\leq 2\,10^{14}$} \\
\hline
\end{tabular}
\mycaption{H$_2$ column densities (in cm$^{-2}$) in the excited (J$\geq$2) levels,
derived 
from a curve of growth analysis by assuming successively b = b$_{\rm CH}$ and 
b = b$_{\rm CH^+}$, 
when available. 
It is likely that the true N(H$_2$) is intermediary or close to one of
these two estimates. 
\label{tab:h2ex}}
\end{center}
\end{table*}
We derive  the high-J 
H$_2$ column densities via a curve of growth analysis by adopting
successively the CH and CH$^+$ b-values. The results are given in 
Table~\ref{tab:h2ex}
and the method is illustrated in Fig.~\ref{fig:growth} in the case of 
the line of 
sight toward HD102065 for which a significantly different b-values have been
measured for CH and CH$^+$.  Fig.~\ref{fig:growth} shows the fit of 
all  measurements for levels J=2 to J=5 to the curve of growth corresponding to
b(CH) (b = 1.9 km$^{-1}$, top) and b(CH$^+$) (b = 3.0 km$^{-1}$, bottom). 
One must keep in mind that 
significant uncertainties can be attached to these column densities in case
the velocity distribution of the excited H$_2$ gas is different from that
of CH and CH$^+$. However in view of the general good linear 
correlation of  N(CH)  with N(H$_2$) at these column densities (Mattila, 1986),
and the correlations of N(CH$^+$) with excited H$_2$ column densities 
(see Sect.~\ref{sec:warm}), it is likely that the J$\geq$2 column densities
are intermediary or close to one of these two estimates.
\section{H$_2$ formation} \label{sec:form}
\subsection{Determination of the $n R$ product}
The local balance between H$_2$ formation and photo-dissociation can be
written as: $ n_{\rm H I}  \, n \, R = n_{\rm H_2} \, \beta_0 \, G \, S  $, 
where 
$ n_{\rm H I}$, $n_{\rm H_2}$, $n$ are the atomic, molecular and total hydrogen
densities ($ n= n_{\rm H I} + 2 \, n_{\rm H_2}$), $R$ the H$_2$ formation rate,
$\beta_0 $ is the 
Solar Neighborhood value of the H$_2$ photo-dissociation rate in the
absence of shielding, $G$ is the radiation field value in
Solar neighborhood units and $S$ a shielding factor including dust
extinction and H$_2$ self-shielding. For the stars studied here, the
IRAS maps show that the stars do not heat the matter responsible for
the absorption, we can thus assume that the stars do not contribute
to the radiation field and  we can take $G$=1 (Boulanger et al. 1994). 
For a constant $n$,
integration over the line of sight leads to: 
$n R = \frac{1}{2} \frac{f}{1-f} \beta_0 \, <S> $
where $f$ is the molecular hydrogen fraction: $f$=2 N(H$_2$)/N$_{\rm total}$, 
where N$_{\rm total}$,
the total column density is derived from the extinction, 
N$_{\rm total}$ = 5.8 10$^{21}$ E(B-V), and $<S> $ is the mean 
shielding factor. 
From this formula, one thus sees
that the product $ n  R $ can be derived from the measured molecular
hydrogen fraction $f$.

Practically a model is necessary to determine the abundance and
distribution of  H$_2$ molecules over its ro-vibrational levels as
a function of depth into the cloud and thereby derive the $n  R $
product from $f$. We have used an updated version  
of the stationary model  developed by Abgrall et al. (1992) and Le Bourlot et
al. (1993).  The model
assumes a semi-infinite plane parallel geometry and solves the equations of 
thermal and chemical balance iteratively as a function of depth into
the cloud. Transfer in the H$_2$ lines (50 ro-vibrational levels are
included in these calculations) and dust extinction 
are taken into account.  
The far UV extinction curves used in the models are specific   
for each star and present a wide variety in terms of 2200 \AA\ bump and 
far UV rise. These curves are extrapolations of  fits 
 to the IUE extinction curves performed by
Boulanger et al. (1994).
The chemical network includes 100 species and 775 reactions. 
To take into account UV penetration from two sides, we run models with
half the total extinction and multiply all integrated column densities
by two. For all calculations, 
we have assumed a constant gas density through the line of sight.
Since we can assume G=1, the model has only two free
parameters, the gas density $n$ and the product $n R$.
The gas kinetic temperature is calculated in the model at each depth 
assuming local equilibrium
between heating and cooling processes. The dominant heating process
is the photo-electric effect on small grains. Heating by dissipation
of turbulent kinetic energy is not included.
\begin{figure}[h]
\ifREFEREE 
  \resizebox{\hsize}{!}{\includegraphics{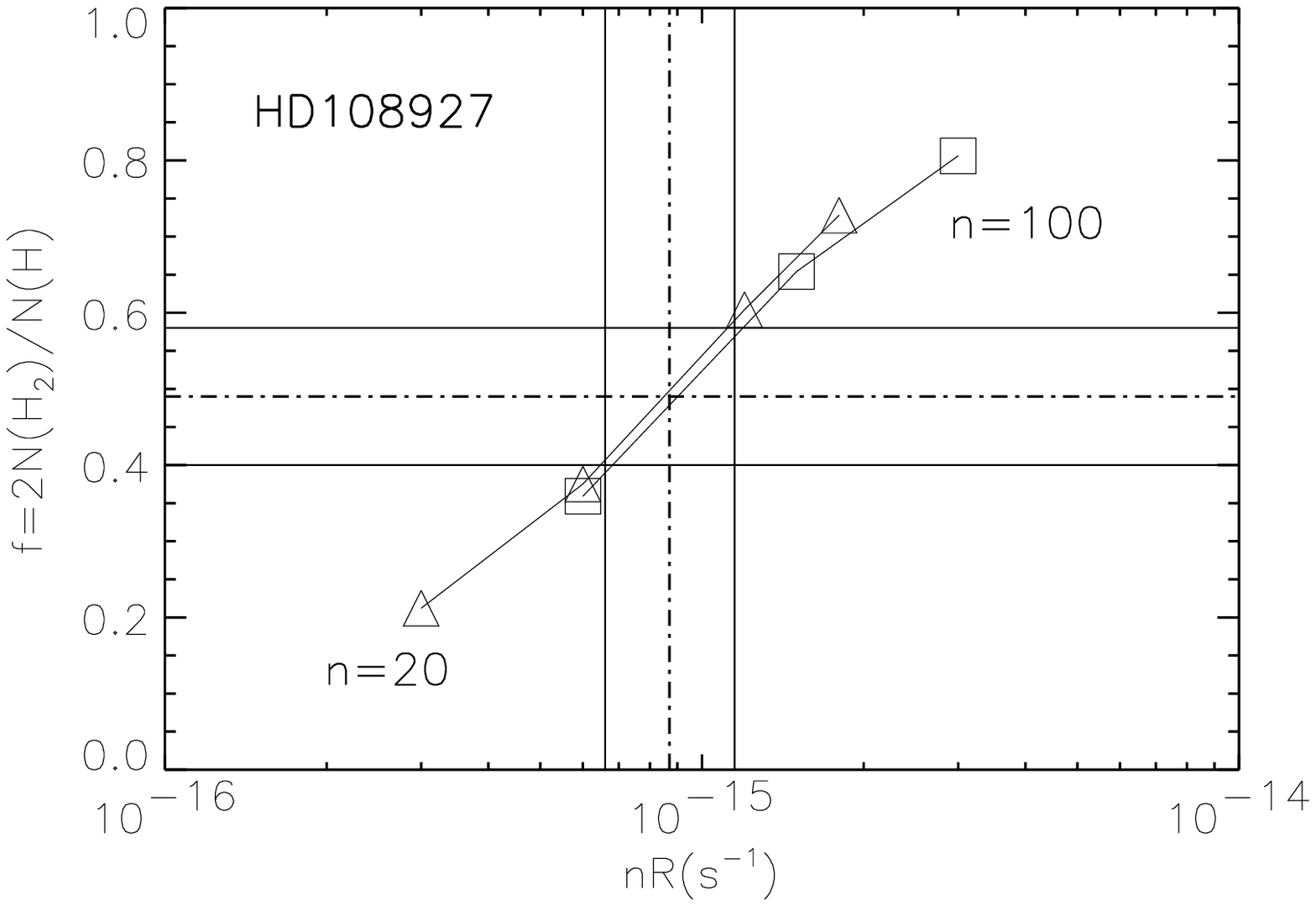}}
\else
\resizebox{\hsize}{!}{\includegraphics*{h3639f3.ps}}
\fi
\mycaption{Determination of the $n R$ product from the H$_2$
fraction $f$ for HD 108927. The triangles and square symbols represent
model calculation for $n$=20 and 100 cm$^{-3}$, respectively. 
\label{fig:form}}
\end{figure}
Model results are shown in Fig.~\ref{fig:form} for the line of
sight towards HD 108927, where the computed H$_2$ fraction $f$ is
plotted versus the product $n R$. A comparison of the  model results 
for n=20 and 100 cm$^{-3}$ illustrates the density independence of the 
relationship. The horizontal lines in the
figure represent the measured $f$ with its error bar 
(from Table~\ref{tab:columns}). The
intersections of these lines with the model results  define the
range for  $n R$. 
We have used the model to quantify the uncertainties on the product 
$n R$
related to the line of sight extinction and the far-UV radiation field.
We find that the error bar on the A$_{\rm V}$ measurement translates into an
uncertainty
of $\pm 50\%$ on $n  R $. An uncertainty of a factor 2 on the far-UV 
radiation field intensity, which is a reasonable assumption, 
translates into an error-bar of the same amplitude on $n  R $.
We thus consider that within the frame work of the model, a factor of
2 is the magnitude of the error-bar on $n  R$. Additional
systematic errors could be coming from the simplifying assumptions made in 
the model. These are of course impossible to quantify. 

The $n R$ values derived for the three stars are
listed  in Table~\ref{tab:formrate}.
Their scatter is of a factor of 2, comparable to the uncertainty on the 
individual measurements. 
Jura (1975b) had found
a much larger scatter with values from  
$5 \, 10^{-16}$ to $ 3 \, 10^{-14}$ s$^{-1}$. 
The dispersion of Jura's values might be partly
due to changes in the G value among the lines of sight towards the
very luminous stars observed by Copernicus. 
\subsection{Determination of R from the H$_2$ density estimate}
To determine the H$_2$ formation rate $R$ one must complement the $n R$
values with an estimate of the gas density $n$.

\begin{figure*}[htb]
\ifREFEREE 
  \resizebox{\hsize}{!}{\includegraphics*{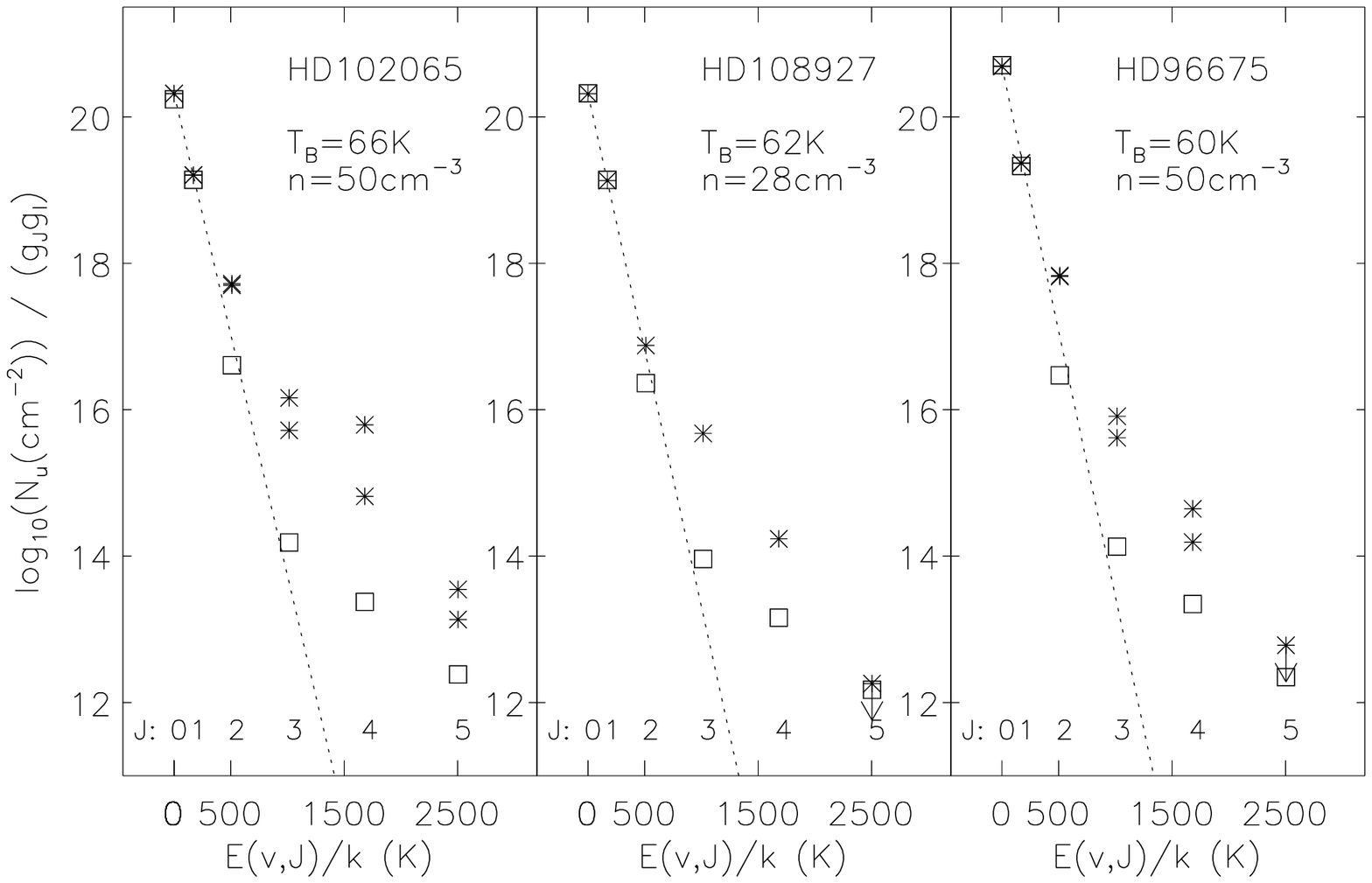}}
\else
\resizebox{16cm}{!}{\includegraphics*{h3639f4.ps}}
\fi
\mycaption{H$_2$ excitation diagrams. The asterisks  represent the measured 
column densities listed in 
Table~\ref{tab:columns} for J=0 and J=1 and in Table~\ref{tab:h2ex} 
for the higher J levels within the two assumptions b(H$_2$)$_{\rm ex}$=b(CH) 
(higher values)
and b(H$_2$)$_{\rm ex}$=b(CH$^+$) (lower values). 
The squares  represent the model that best fits the J=0 and J=1 observations
and corresponds to the density indicated. 
The dashed line is the 
Boltzmann distribution for the temperature $T_{\rm B}$ corresponding to the observed 
N(J=1)/N(J=0) ratio. 
\label{fig:h2-excitation}}
\end{figure*}

The gas density $n$ is the second free parameter of the model.
Density governs the cooling rate and the assumption of thermal balance makes 
the gas density and
temperature uniquely related at each depth
once the external radiation field is given. 
Any tracer of the gas temperature is therefore  also a tracer
of the density. We thus determine the gas density by fitting the model to
the column density ratio of the two first levels, N(J=1)/N(J=0), known
to be an indicator of temperature.
 Fig.~\ref{fig:h2-excitation} 
illustrates the best fits, corresponding to the density values indicated 
in each diagram. Note that in the model the temperature is not uniform, 
its value is determined at each depth by the thermal balance. The temperature 
$T_{\rm B}$ indicated in Fig.~\ref{fig:h2-excitation}  is that of the shown 
Boltzmann 
distribution  corresponding to the N(J=1)/N(J=0) ratio value.

The  $R$ values derived for the three lines of sight, based on these density 
estimates are listed in Table~\ref{tab:formrate}. 
This determination of $n$ assumes that the ortho to para H$_2$ ratio is at its 
equilibrium value and is therefore very sensitive to the temperature, which
is itself, within the assumption of thermal balance, very sensitive to
the gas density. Within the model framework,
the H$_2$ J=1 and 0 column density ratio constrains density within about
20 \%, an uncertainty smaller than that on the product $n R$. 
The uncertainty on $R$ is thus governed
by the uncertainty on $n R$: about a factor of 2. 
 
The values of $R$ for the three lines of sight are  close to each other 
and close to the 
values found by Jura (1975a) but with  significantly  lower uncertainties.
\begin{table}[h] 
\begin{flushleft}
\begin{center}
\leavevmode
\footnotesize
\begin{tabular}{c c c c}
\hline 
   Etoile     & HD 102065 & HD 108927 & HD 96675 \\
\hline
	\noalign{\smallskip}
$n R$ (s$^{-1}$) & 2.3 10$^{-15}$ & 0.87 10$^{-15}$ & 2.0 10$^{-15}$ \\       
$n$(cm$^{-3}$) & 50 & 28  & 50 \\
$R$(cm$^3$s$^{-1}$) & 4.5 10$^{-17}$ & 3.1 10$^{-17}$ & 4.0 10$^{-17}$ \\
\hline
\end{tabular}
\caption{H$_2$ formation rate $R$ from the product 
$n R$ and the density $n$ estimated from N(H$_2$,J=1)/N(H$_2$,J=0).
\label{tab:formrate}} 
\end{center}
\end{flushleft}
\end{table}
Note nevertheless that the derivation of $R$ from $nR$ is valid within 
the hypothesis that the density is homogeneous in the molecular gas. 
This might not be the case. Indeed, for HD
102065 for which we have been able to derive three other independent density 
estimates from the comparison of model
calculations and measured quantities (namely N(C\,{\sc i}), N(CH) and the 
C\,{\sc i} fine 
structure level population from Gry et al. (1998)), there is a scatter
of a factor of three among the four density estimates. 
 This scatter might well reflect a density
inhomogeneity along the line of sight but could also result from
other model shortcomings, in particular in our understanding of the
chemistry in diffuse clouds.

The $n R$ values directly translate into an H$_2$ formation timescale
of  $1/(n  R) \sim 2 \, 10^7$yrs.  For the three stars it is a few times
 larger than the dynamical
timescale ($\rm \sim  10^6 L(pc)/b(km/s) yrs$) set by turbulent motions 
on the scales of the absorbing
clouds ($\rm L={N}/{\it n} \sim 10 pc$). The model tells us that the 
photodissociation timescale is larger than the H$_2$ formation timescale
especially in the shielded layers of the cloud (the dissociation timescale 
goes from about $3 \, 10^7$yrs to about $3 \, 10^8$yrs for A$_{\rm V}$ from 0.3 
to 1). On the other hand the timescale of ortho to para conversion through 
proton exchange reactions with protonated ions (e.g. H$^+$ and H$_3^+$, 
Gerlich 1990)
is much shorter (0.2 to 2 $10^5$yrs). 
Consequently, the
observed H$_2$ abundance may not correspond to the equilibrium value between
formation and destruction and
by assuming equilibrium in our interpretation we could be overestimating 
$R$ and the $n R$ product. 
Study of a larger sample of stars could reveal scatter in the $R $
and $n R$ values reflecting various states of evolution. 
\section{Warm H$_2$ gas} \label{sec:warm}
The excitation diagrams of Fig.~\ref{fig:h2-excitation} show
that the H$_2$ column densities derived from the observations for $\rm J>2$ 
are all significantly  higher  than
the model values. This is true for our two determinations of H$_2$ column 
densities, the
highest based on the Doppler parameter of CH and the lowest based on that of 
CH$^+$ (see Sect.~\ref{sec:h2col}). This means that H$_2$ excitation by 
UV pumping and H$_2$ formation on grains as computed with the model
does not account for the $\rm J>2$ H$_2$ column densities.
With the model, we checked on HD 102065 that to populate these levels  
by UV pumping alone (with no collisional deexcitation) 
we would have to increase
the radiation field to a G value of 17 to reproduce the column densities
derived with b(CH) and to a G value of 9 to reproduce the column densities
derived with b(CH$^+$). Such  values are clearly
incompatible with the IRAS dust data because no enhanced IR dust emission
is found at the position of the stars and the ratio between the $100 \mu$m 
cloud brightness and extinction is within the range of values
observed over the high Galactic sky. 
We have checked also that  we cannot 
reproduce the measured J=2 level populations
even if all the H$_2$ binding energy (4.48 eV) is  transformed 
in H$_2$ internal excitation.  

\begin{table*}[t] 
\begin{flushleft}
\begin{center}
\leavevmode
\footnotesize
\begin{tabular}{l c c c c}
\hline 
        & HD102065 & HD108927 & HD96675 & ISO \\
	\hline
	\noalign{\smallskip}
	N(H$_2$)$_{\rm ex}$  & $1.7-3.7 \, 10^{17}$ & $1.0 \, 10^{17}$ & $0.9-1.7 \, 10^{17}$ & $7.7 \, 10^{18}$ \\
	A$_{\rm V}$   & 0.67 & 0.68 & 1.1 & 18 \\ 
	$\rm \frac{N(H_2)_{\rm ex}}{A_{\rm V}}$  &$2.5-5.4 \, 10^{17}$ & $1.5 \, 10^{17}$ & $0.8-1.6 \, 10^{17}$ & $4.3  \, 10^{17}$  \\
	\noalign{\smallskip}
	\hline
	\end{tabular}
	\mycaption{Abundance of excited H$_2$ gas.       
	N(H$_2$)$_{\rm ex}$ (in cm$^{-2}$) is the sum of the J=3 to 5 column 
	densities from Table~\ref{tab:columns}. \\
The last column (ISO) refers to an infrared observation of a  long line of 
sight through the Galaxy (from Falgarone et al. 2002).  
	\label{tab:warm}}
	\end{center}
	\end{flushleft}
	\end{table*}

We thus infer the existence of warm gas along the three lines of sight, 
where the  J$>$2 H$_2$ levels are populated by collisional excitation. 
This statement relies on the validity of the H$_2$ column densities
determinations. In the absence of any direct information, one cannot exclude 
that 
the H$_2$ velocity distribution is broader than
that of both CH and CH$^+$ and consequently that even the lowest set of values 
in Table~\ref{tab:h2ex} are larger than  the true  H$_2$ column densities.  
However, we consider unlikely that the H$_2$ velocity distribution could be such 
that there is no need for warm H$_2$ gas because the detection of CH$^+$ in the
direction of both HD 96675 and HD102065  where it has been looked for 
is an independent evidence for
the existence of a warm H$_2$ component along these lines of sight. 
If one considers only collisional excitation, 
the H$_2$ excitation temperature is  
lower than the true gas temperature for densities below the critical densities of 
each level.  The temperature of the warm gas thus needs to be at least equal to the 
H$_2$ excitation temperature derived from the J=3 to 5 levels, i.e. between
200 and 240 K for the three stars.

Our conclusion about H$_2$ excitation at J$>$2 levels
differs from  a common interpretation of UV
H$_2$ absorption lines (e.g. Jura 1975b) where it is assumed that the high J 
levels are mainly populated by pumping through UV photons of the target stars.
For our sample of stars  we are able to rule out this interpretation, because  the 
FUSE sensitivity  allowed us to observe late B stars which do not interact 
with the absorbing matter, thus lines of sight for which the UV 
radiation field strength is constrained to be close to the mean Solar Neighborhood value.  

The existence of warm H$_2$ gas within the diffuse ISM has been considered along many 
lines of sight to account for the observed column densities of CH$^+$.
The observed abundance of CH$^+$ is a well known problem of interstellar chemistry.
The only efficient path for CH$^+$ formation is the highly
endothermic (4640 K) reaction between C$^+$ and H$_2$. Further, CH$^+$ is
efficiently destroyed by reaction with H$_2$ once it is formed. 
One thus considers that CH$^+$ only exists in significant abundance where the 
molecular gas is warm. Away from hot stars, localized volumes of warm gas 
can be created and sustained  by dissipation of the gas kinetic energy. 
Formation of CH$^+$ have 
been quantitatively investigated in the specific cases where dissipation occurs within  MHD 
shocks (e.g. Flower \& Pineau des
For\^ets, 1998) or coherent vortices
in MHD turbulence (Joulain et al., 1998). In these models the temperature of the warm
gas and the ratio between CH$^+$ and warm H$_2$ column densities depends strongly on
local physical conditions (e.g. shock velocity, gas density, magnetic field 
value). The addition 
of H$_2$ excitation studies provides a mean to constrain these models and 
can thus help understand the physics of 
kinetic energy dissipation in the diffuse ISM. 

For each line of sight we have computed and listed in Table~\ref{tab:warm} 
a total column density of excited H$_2$ gas (N(H$_2$)$_{\rm ex}$) by
summing the column densities in the J=3 to 5 levels. The range of values
corresponds to the two assumptions used for the determination of H$_2$ column 
densities (see Sect.~\ref{sec:h2col}). The fact that CH$^+$ must form in warm H$_2$ gas makes
us consider the lowest value of (N(H$_2$)$_{\rm ex}$, those obtained for
the CH$^+$ Doppler parameter, to be the most realistic estimate. For HD 108927, 
in the absence 
of CH$^+$ observation only
the higher estimate is listed.  
Note that in all cases the excited H$_2$ gas represents a very small fraction
of the total gas.
We have not attempted to separate
the respective contributions of UV pumping in the cold gas and truly warm gas 
to the column densities of excited H$_2$.
The excitation diagrams in Fig.~\ref{fig:h2-excitation} show that UV pumping 
is always a minor contribution. 

Based on Copernicus data, Frisch \& Jura (1980)
suggested a correlation between the  column densities of H$_2$
in the J=5 level and  CH$^+$. They 
pointed out that CH$^+$ is particularly abundant in regions with large amounts of
rotationally excited H$_2$. This result has been confirmed by 
Lambert \& Danks (1986) who compared the column densities N(CH$^+$) and 
N(H$_2$,J) for the rotational levels J=0-5 in a larger sample of sight-lines. 
They concluded  that the 
rotationally excited H$_2$ is a tracer for those hot H$_2$ molecules that 
initiate CH$^+$ formation. Both lines of sight of our study, HD102065 and HD96675,
for which the CH$^+$ column density is known,  fit within the
dispersion of the Lambert \& Danks (1986) correlation diagrams. Note that the
Lambert \& Danks' sample consist of luminous stars for which one cannot exclude that
the excited H$_2$  gas results from the interaction (UV pumping and shock) 
of the star with its surrounding medium. Further sight-lines such as those
considered here, for
which this possibility can be excluded, are needed to physically discuss  the 
correlation between CH$^+$ column densities and warm H$_2$ in relation with 
existing models of CH$^+$ formation. 

One Galactic line of sight not crossing any known star forming region
has been observed with the mid-IR spectrometer on board of the
Infrared Space Observatory (ISO). This observation lead to the
detection of three  pure rotational lines of the H$_2$ ground state
(Verstraete et al. 1999, Falgarone et al. 2002). The H$_2$ populations 
inferred from these
detections are compared with our FUSE results in Table~\ref{tab:warm}.  The 
N(H$_2$)$_{\rm ex}$/A$_{\rm V}$ ratio for the long line of sight through the Galaxy is  
a factor 5 and 2 higher that  the lower estimates 
listed  for HD 96675 and HD 102065. We thus find along these two sight-lines 
a fraction of warm H$_2$ gas close
to the mean Galactic value derived from the ISO observation. The presence of warm H$_2$
thus seems to be a general characteristics of the diffuse ISM. 

 \begin{acknowledgements} 
We are grateful to Martin Lemoine for  his profile fitting software
Owens and to Vincent Lebrun for reprocessing the data with the  FUSE
pipeline version 1.8.7.
This work is based on data obtained for the Guaranteed Time Team by the
NASA-CNES-CSA FUSE mission operated by the Johns Hopkins University.
\end{acknowledgements}

\end{document}